# Shannon entropies of the distributions of various electroencephalograms from epileptic humans

Çağlar Tuncay

Department of Physics, Middle East Technical University

06531 Ankara, Turkey

caglart@metu.edu.tr

**Abstract**: In this letter, nearly 700 million data recorded from nearly 20 epileptic humans with different brain origins of epilepsy, ages or sexes are analyzed, and;

1) Harmonic oscillations (HO) in numerous electroencephalograms (EEG) from different humans are introduced.

Inspection of the data shows that HO may come out besides the ordinary ones (OO), for several seconds or hours or longer in several simultaneous individual recordings from different brain sites in an inter-ictal interval or ictal interval. HO are deformed in certain time intervals (epoch) when the cyclic behavior is altered or wave amplitude is time dependent. Then the individual oscillations become mixed (MO). Thus, the EEG oscillations can be categorized mainly in three groups; HO, OO or MO.

2) The probability density functions (PDF, $p(X)$) of the EEG voltages (X) are normal (Gauss) for OO whereas, the plots for the distributions of HO (pure) are convex. Gaussians for OO may turn to be convex as HO become dominant in MO or vice versa. However, distributions of the most of the data are found normal which means that most of the EEG oscillations consist of OO (or MO).

3) Shannon entropies (information measures) of the distributions of the data from different brain regions in the ictal intervals or inter-ictal intervals are calculated for each individual recording and compared. The averages of Shannon entropies over the individual recordings during the ictal intervals come out bigger than those from the inter-ictal intervals. These averages are found to be bigger for the data from epileptogenic brain areas than those recorded from non epileptogenic ones in different intervals.

**Pacs**: 87.19.Nn, 87.15.Aa, 05.90.−y; 05.90.+m; 87.10.+e; 87.59.Bh

**Introduction**: Patterns of the EEG signals are widely studied in various linear or non linear based approaches [1]. These analyses may be valuable for detection or prediction of epileptic seizures as pointed in [2]. Another suggestion is that EEG voltages with big absolute value (amplitude) [3] maybe precursors of epileptic seizure onsets where the number of the big amplitudes may also be important. Thus, entropies of the EEG distributions can be useful for characterizing the EEG data [4]. With this aim, EEG distributions and their Shannon entropies (S) are considered in this letter. [5]

*EEG distributions and Shannon entropies*: Several statistical properties of EEG data are known to be investigated in terms of their distributions, from the 1950s on. [6] A recent treatment of human EEG distributions may be found in [7]. Entropies of the distributions are also studied in various contexts (for direct applications, see [8]).

Entropy is known to be a thermodynamic quantity describing the amount of disorder in a system. It can be taken as a measure of uncertainty in the information content. Shannon entropy is the measure used to analyze human EEG signals in this letter. (For entropies in EEG data from animals see, [8] or references given therein.)

If P(X) is a normalized distribution of the brain voltages (X which are integers in micro Volts (µV), here), then S is

$$S = -K_B \sum_i P_i \ln(P_i) \qquad (1)$$

where the summation is over the states (i) which are accessible with probability ($P_i$), ln is the natural logarithm and $K_B$ is Boltzmann constant which is treated as unity and the equality sign is replaced by $\propto$, here.

S in Eq. (1) can be related to the standard deviation ($\sigma$) or height ($p_{max}$) of a normal distribution (p(X)) about a mean ($\lambda$);

$$p(X) = (2\pi\sigma^2)^{-\frac{1}{2}} \exp(-(x-\lambda)^2/2\sigma^2) ; \qquad (2)$$

$$S \propto \tfrac{1}{2}(\ln(2\pi\sigma^2) + 1) = \ln\sigma + 1.4189 \qquad (3)$$

or

$$S \propto \tfrac{1}{2} - \ln(p_{max}) \quad , \qquad (4)$$

respectively. The summation in Eq. (1) is approximated by integration for the results given in Eqs. (3) or (4) (or (6), below).

Normalized distributions ($p_{HO}$) of HO about a mean ($\lambda$) follow;

$$p_{HO}(X) = \pi^{-1}(Q^2-(X-\lambda)^2)^{-\frac{1}{2}} \quad \text{for } -Q<X<Q \qquad (5)$$

where Q is the wave amplitude which may be constant or time dependent in different epochs in the recordings of a person or different persons. HO can be shown to have the following entropies for various Q values:

$$S_{HO} \propto \ln Q + 0.45158 \qquad (6)$$

Note that PDF in the Eqs. (2) or (5) are concave or convex for OO or HO, all respectively. If the oscillations are mixed, then the tops of the distribution peaks may come out concave or flat, depending on the relative amount of HO in the data. Secondly, entropies (Eq. (1)) are big for normal distributions (Eq. (2)) with big standard deviations (Eq. (3)) or small heights (Eq. (4)) and similarly for convex distributions (Eq. (5)) with big wave amplitudes (Eq. (6)), where the data may be considered as dispersed over the voltages in all.

*Materials and Methods*:    A pool of data [9] recorded directly from the brain tissues of 21 epileptic patients in inter-ictal interval or ictal interval is investigated. The ages or sexes of the patients are different. The data from few patients, namely the Pat02, Pat07, Pat09 and Pat13 seem problematic and thus they are disregarded in the present analysis. The brain origins of the epilepsies of the considered patients are parietal (Pat11), frontal (Pat01, Pat03, Pat05, Pat08, Pat18 or Pat19), fronto-temporal (Pat14), temporal (Pat04, Pat10, Pat12, Pat15-17 or Pat21), temporo-occipital (Pat06) or tempo-parietal (Pat20).

The registrations are performed by 6 electrodes, simultaneously; the first three of which are in-focus (k=1-3) and the rest are out-focus (k=4-6). The rate of the recordings (f) of the considered data is 256 Hertz. Numerous non linear based studies on the same data set are found in [10].

Here linear analysis is followed:
1) Distributions of the data recorded from each considered patient (j=1-17) using the electrode connected to the epileptogenic brain areas (k=1-3) or non epileptogenic ones (k=4-6) in ictal interval (m=1) or interictal interval (m=2) are computed and normalized ($P_{jkm}$).
2) Gauss the least square fits (LSF) are applied to $P_{jkm}$ and thus the means ($\lambda_{jkm}$) or standard deviations ($\sigma_{jkm}$) are estimated. Besides, the degrees of goodness of the fits are given in terms of the coefficient of determination (COD, $R^2_{jkm}$).
3) The individual Shannon entropies ($S_{jkm}$) are calculated;

$$S_{jkm} \propto - \sum_i P_{jkm,i} \ln(P_{jkm,i}) \qquad (7)$$

where ($P_{jkm,i}$) designate different brain states with k, m or i as before and j=1-17 for the Pat01, Pat03-Pat06, Pat08, Pat10-Pat12, Pat14-Pat21, respectively. The means, standard deviations or entropies are compared with respect to numerous criteria. The results are presented in the following section. The last section is devoted to discussion and conclusion.

**Results**:

*Harmonic, ordinary or mixed oscillations*: Figure 1 (a) illustrates a segment of a time chart showing HO recorded from the Pat04 (j=3) with temporal epilepsy during ictal interval by an in-focus electrode (k=1). Note that the pulses in Fig. 1 (a) are seen as a train of pinlike (triangular) pulses when the data points are connected by a line. Figure 1 (b) is the same as Fig. 1 (a) but with a time domain of one hour (9,216,000 data points recorded with f=256 Hertz) where the time variation of the wave amplitude (Q in Eq. (5)) is obvious. The wave amplitudes (Q) of HO from the Pat04 are constant in discrete epochs but they may abruptly change from the end of an epoch to the beginning of the consecutive one while several relaxations or transitions (spiky behavior) take place in or between the mentioned epochs, respectively. The means also vary from one epoch to another (not shown). Hence, the convex topographies for each epoch overlap with each other and these with big Q become unseen in the related PDF plots. The PDF of all of the data recorded from an epileptogenic site of the Pat04 by the electrode (k=1) is given in (c) which is not Gauss. Note the time period of HO recorded from the Pat04 is little less than 1 second as exemplified in Fig. 1 (a) where the registration rate is 256 Hertz.

HO emerge also in the recordings from different brain sites (k=2-4) of the Pat04 throughout the ictal interval and their distributions are convex as those shown in Fig. 1 (b). The data in the individual recordings (k=5, 6) from the ictal interval or all of the data recorded from several brain sites (k=1-6) of the same patient during the inter-ictal interval distribute normal (concave). Hence, these oscillations are either OO or MO.

HO can be inspected also in the EEG data recorded from epileptogenic or non epileptogenic brain areas of numerous patients in different intervals where the distributions may be normal, not convex. This is because, these HO may have time dependent Q or λ. Thus, HO are modulated and they are detected as MO. For example, the EEG data from the Pat03 with frontal epilepsy recorded by the first electrode (in-focus) during the ictal interval involve HO in several epochs. But, they are not clearly seen in the time charts as exemplified in Figure 2 (a).

The means varying linearly with the time may be eliminated in terms of the successive differences of the voltages ($\Delta X(n)$);

$$\Delta X(n) = X(n+1) - X(n) \quad . \tag{8}$$

Figure 2 (b) shows the same segment of the time series as in Fig. 2 (a) except for the fact that the recently mentioned means are removed. HO are clear in Fig. 2 (b).

Also several data segments from the Pat16 with temporal epilepsy recorded during the ictal interval involve HO (not shown) in several epochs. These HO may be inspected more clearly in the time series for $\Delta(\Delta X(n)) = \Delta(X(n+1)) - \Delta(X(n))$ than those for $\Delta X(n)$. In this case, the means vary not linearly but in a complicated manner. Note that all of the distributions (PDF) of all of the individual recordings from the pats Pat03 or Pat16 are normal with various degrees of goodness.

Figures 3 (a) or (b) are for OO recorded from the Pat04 during the ictal interval but from a non epileptogenic brain site (k=5) or the related distributions (of 9,216,000 data), respectively. The plot for PDF in Fig. 2 (b) is concave for the small voltage magnitudes.

The tails of both, the concave or convex distributions in Figs. 1 (c) or 3 (b) are contaminated by the relaxation or transient voltages, or spikes with big magnitude.

*Gauss LSF*: Several parameters for the applied Gauss LSF to the distributions of the individual recordings (200 peaks for MO or OO) are displayed in the Figures 4 (a)-(c) where the errors in the estimations $\lambda_{jkm}$ or $\sigma_{jkm}$ are less than 1.0 μV.

The related results may be summarized as: 1) The most of the experienced LSF may be considered as good since the parameters for COD ($R^2_{jkm}$) are bigger than 0.9 where the average equals nearly 0.98 (Fig. 4 (a)). 2) The means are different and they show nearly symmetric variations about the zero if they are in ascending (or descending order) where the overall average is nearly 0.011 mV. 3) Nearly half of the standard deviations are bigger than their average value which is about 0.66 mV where the maximum value equals nearly 2 mV. Hence, 99.7 % of the data in each peak lie within nearly a few mV far from the mean (confidence interval) following the well known "three standard deviation rule". It can be claimed that the distributions of most of the considered EEG data are normal. Only the distributions of the individual recordings ($P_{jkm}$) from the Pat04 (j=3) by the electrodes (k=1-4) in ictal interval (m=1) are convex since the raw data are HO.

*Shannon entropies*: The individual Shannon entropies ($S_{jkm}$ in Eq. (7)) of the used data [9] show fluctuations as displayed in Figure 5 (a). There, each successive group of 12 points are about one person (j=1-17). The first subgroup of 6 data points in each group is from ictal interval of each patient (m=1) and the second subgroup of 6 points are with m=2 (ictal interval). Furthermore, the solid boxes are for the recordings from epileptogenic areas of the brains and empty circles are for those from non epileptogenic ones. The $S_{jkm}$ for the data from epileptogenic areas are bigger than those from the non epileptogenic ones, except few cases.

Figures 5 (b) or (c) show the averaged entropies over several k; ($<S'>_{jm}$ or $<S''>_{jm}$), or ($<S>_{jm}$) or, respectively;

$$<S'>_{jm} = \sum_{k=1}^{k=3} S_{jkm}/3 \tag{9}$$

or $$<S''>_{jm} = \sum_{k=4}^{k=6} S_{jkm}/3 \tag{10}$$

or $$<S>_{jm} = \sum_{k=1}^{k=6} S_{jkm}/6 \tag{11}$$

where j or m as before. $<S>_{jm}$ in Eq. (11) is the average entropy calculated using the entropies of the individual recordings from different sites. $<S'>_{jm}$ or $<S''>_{jm}$ are for those from the epileptogenic or non epileptogenic brain areas, respectively. Hence, $<S>_{jm}$ are the averages of $<S'>_{jm}$ and $<S''>_{jm}$.

**Discussion and Conclusion**: Apparently, the EEG data involve more information which should be better recognized in medicine or literature. For example, various EEG data from numerous epileptic humans show HO but they are not considered in electroencephalography, in the past or present. Moreover, the wave amplitudes of HO or the means of HO, OO or MO may be different in different epochs where σ also show variations.

Entropies of concave or convex distributions have similar analytical expressions; Eqs. (3) or (6), respectively:

$$S \propto \ln W + Z \tag{12}$$

where W has the dimension of Voltage and Z is a real constant which is different for different types of distributions. Thus, Z indicates the type, and W stands for a characteristic voltage of a distribution for HO, MO or OO:

$$W = \exp(S - Z) = z\exp(S) \tag{13}$$

where z=exp(-Z) and W may be utilized as a measure of how dispersive the data are in a given distribution. Note that W=σ for OO or MO and W=Q for HO with Z=1. 4189 or Z=0.45158, respectively. As a result, big entropies come out for big W (=σ or Q) in Figs. 5 (a)-(c).

**ACKNOWLEDGEMENT**

The author is thankful to the University of Freiburg for their kindness in giving permission to investigate their databases.

**FIGURES**

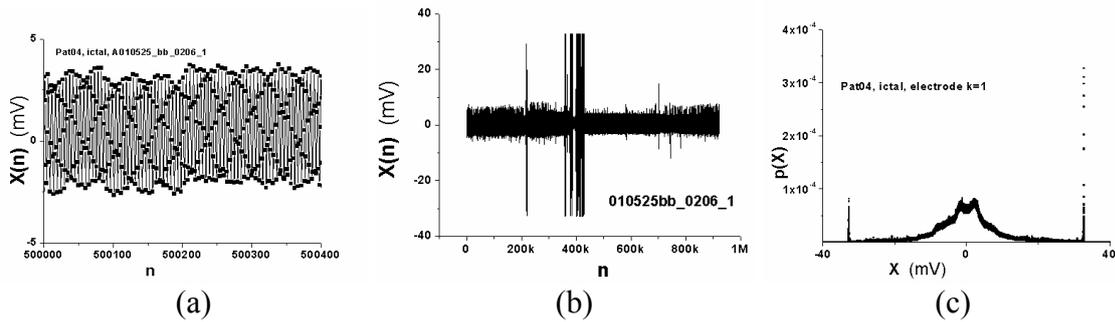

        (a)                                     (b)                                   (c)

**Figure 1**    (a) shows a segment of HO in the time series X(n) of the declared data [9] recorded during ictal interval by the first electrode which is in-focus (k=1) and (b) is the same as (a), where the time domains are nearly 1.5 second or exactly 1 hour, respectively. The PDF of all of the data recorded from epileptogenic areas of the Pat04 by the electrode (k=1) is given in (c) which is not Gauss.

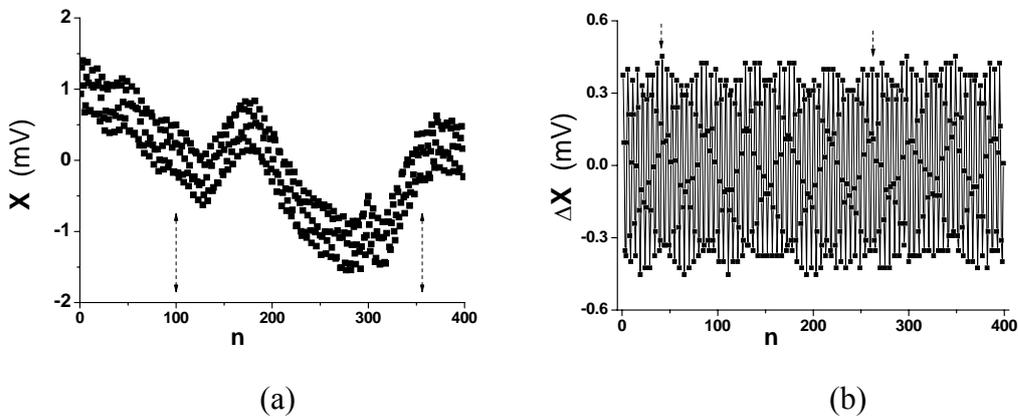

                      (a)                                                            (b)

**Figure 2**    (a) or (b) are the segments of the time charts X(t) (MO) or ΔX(t) (HO), respectively, of the EEG recorded from the Pat03 by the electrode k=1 in the ictal interval. The dashed arrows indicate 1 second or a period in (a) or (b), respectively.

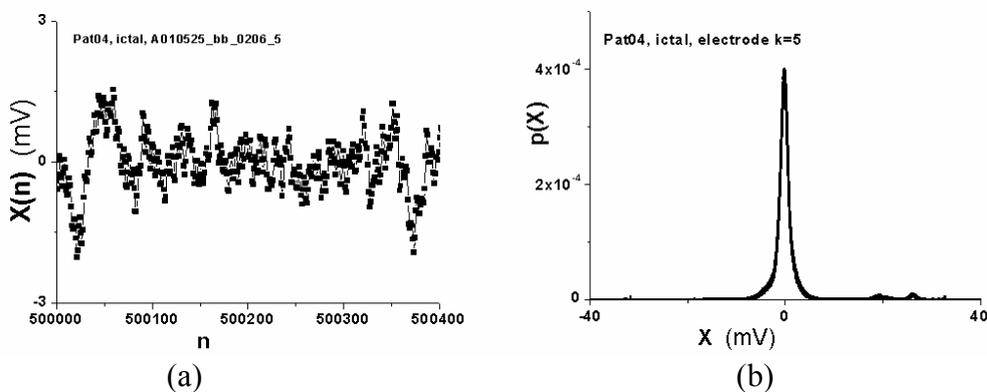

                      (a)                                                            (b)

**Figure 3**    (a) or (b) are the same as the Figs. 1 (a) or (c) but for the data showing OO which are recorded from the fifth electrode (out-focus) and PDF is normal.

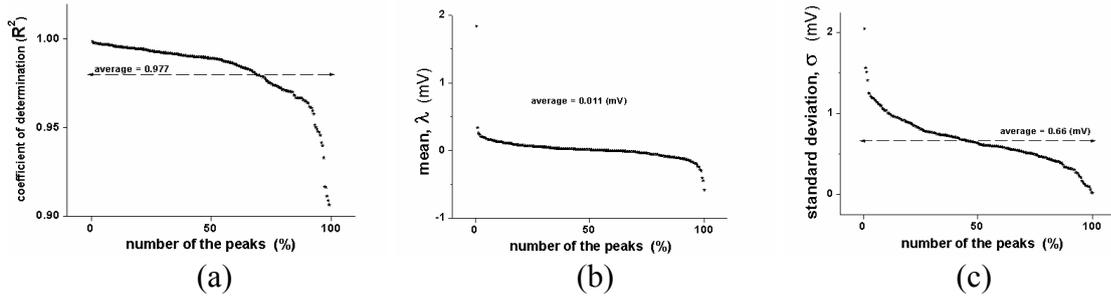

(a)                           (b)                           (c)

**Figure 4**      COD, means ($\lambda$) or standard deviations ($\sigma$) for PDF of the used data are shown in (a), (b) or (c), respectively. The horizontal dashed arrows indicate the parameters for the related averages in (a) or (c).

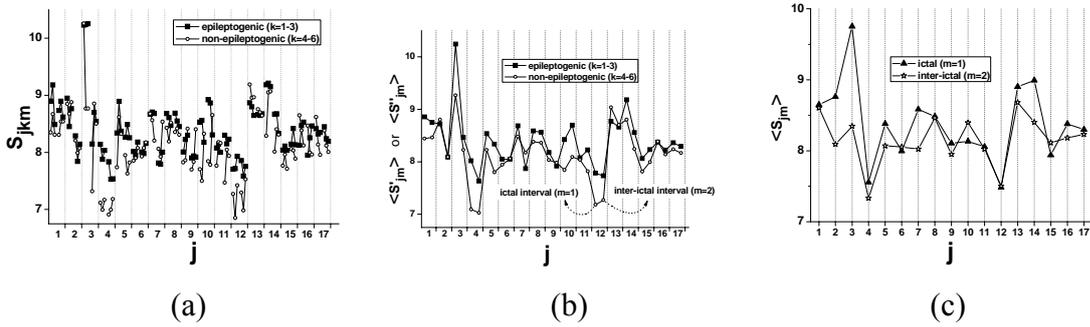

(a)                           (b)                           (c)

**Figure 5**      (a) shows the individual Shannon entropies, and (b) or (c) are for the averages; $<S'>_{jm}$ or $<S''>_{jm}$, or ($<S>_{jm}$), respectively. (See, Eqs. (9)-(11).)